\documentclass[11pt,twoside]{article}
\usepackage{./asp2014}

\aspSuppressVolSlug
\resetcounters

\begin{document}

\title{Collision between Neutron Stars and Asteroids as a Mechanism for Fast Radio Bursts}
\author{Y. F. Huang,$^{1, 2}$ and J. J. Geng$^{1, 2}$
\affil{$^1$School of Astronomy and Space Science, Nanjing University, Nanjing 210046, China;  \email{hyf@nju.edu.cn}}
\affil{$^2$Key Laboratory of Modern Astronomy and Astrophysics (Nanjing University), Ministry of Education, Nanjing 210046, China}}

\paperauthor{Sample~Author1}{Author1Email@email.edu}{ORCID_Or_Blank}{Author1 Institution}{Author1 Department}{City}{State/Province}{Postal Code}{Country}
\paperauthor{Sample~Author2}{Author2Email@email.edu}{ORCID_Or_Blank}{Author2 Institution}{Author2 Department}{City}{State/Province}{Postal Code}{Country}
\paperauthor{Sample~Author3}{Author3Email@email.edu}{ORCID_Or_Blank}{Author3 Institution}{Author3 Department}{City}{State/Province}{Postal Code}{Country}

\begin{abstract}
As a new kind of radio transient sources detected at $\sim 1.4$ GHz, fast radio
bursts are specially characterized by their short durations and high intensities.
Although only ten events are detected so far, fast radio bursts may actually frequently
happen at a rate of $\sim 10^{3}$ --- $10^4~\rm{sky}^{-1}~\rm{day}^{-1}$. We suggest
that fast radio bursts can be produced by the collisions between neutron stars and
asteroids. This model can naturally explain the millisecond duration of fast radio
bursts. The energetics and event rate can also be safely accounted for. Fast radio
bursts thus may be one side of the multifaces of the neutron star-small body collision events,
which are previously expected to lead to X-ray/gamma-ray bursts or glitch/anti-glitches.
\end{abstract}

\section{Introduction}

Fast radio bursts (FRBs) are short bursts of radio emission from the sky. They were discovered
recently (Lorimer et al. 2007; Thornton et al. 2013), with about only ten events being reported
as definite detections till July 2015 (Keane et al. 2011; Burke-Spolaor \& Bannister 2014; Spitler
et al. 2014; Ravi et al. 2015; Keane \& Petroff 2015; Maoz et al. 2015; Connor et al. 2015).
They are usually detected at $\sim 1.4$ GHz, by the large radio telescopes such as the 64-m
Parkes radio telescope and the 305-m Arecibo telescope.  These events are of extremely short
duration, typically lasting for less than a few milliseconds, but are detected with high intensity,
usually with a peak flux density of nearly $\sim 1$ Jy. A terrestrial origin for them
are safely excluded for several reasons. First, the radio telescopes are uniformly pointing toward the
sky at the time of the detections. Second, for the multi-beam receiver system, usually the
signal was recorded only in very few beams (typically less than 4, especially by adjacent beams).
Third, FRBs are characterized by large dispersion measure (DM) values, significantly larger
than terrestrial sources of interference. Fourth, the observed behaviors of time delay and
frequency evolution of FRBs strongly indicate that cold-plasma dispersion should have been engraved
in the radio signal.

FRBs are generally screened out from archive data, thus cannot be quickly followed up to catch
the counterparts in other wavelengths. Recently, a nearly real-time FRB event (FRB 140514) triggered
an extensive search for the counterpart by multi-wavelength telescopes within hours of the event,
but still led to null results in X-ray, optical, and radio bands (Petroff et al. 2015). The absence
of counterparts leads to great difficulties in understanding the nature of these enigma bursts.
From the large DM values, many authors deduced that FRBs should be at cosmological distances,
typically with redshift of $0.1 < z < 1$. The characteristic radio luminosity ($L_{\rm FRB}$)
is then $10^{42}$ --- $10^{43} ~\rm{erg}~\rm{s}^{-1}$ and the isotropic energy release
($E_{\rm FRB}$) will be $10^{39}$ --- $10^{40}~\rm{erg}$. Also, the quick variability of FRBs
indicates that the emission region should be very small. Based on these facts, various models
have been suggested. Some authors argued that FRBs are connected with the giant flares of
magnetars (Popov \& Postnov 2007; Lyubarsky 2014;  Kulkarni et al. 2014;  Pen \& Connor 2015),
others believed that they correspond to the collapse of hypermassive neutron stars into
black holes (Zhang 2014; Falcke \& Rezzolla 2014; Ravi \& Lasky 2014). They were also
suggested to be caused by planetary companions around pulsars (Mottez \& Zarka 2014) or
double neutron star mergers (Totani 2013). Other authors even proposed that FRBs come from
flare stars (Loeb et al. 2014), binary white dwarf mergers (Kashiyama et al. 2013), or
evaporation of small black holes(Barrau et al. 2014).
In many of these models, a very strong electro-magnetic outburst (something like a gamma-ray
burst) would be triggered and should be observed to associate with the FRB event. Multi-band
afterglows are also expected after the FRBs. However, these phenomena have not been observed.
Also, some of the models have difficulties in accounting for the event rate of FRBs.

Recently, we proposed a new model for FRBs (Geng \& Huang 2015). We argued that they
could be produced by the collision of asteroids with neutron stars. The model
can account for many of the observational characteristics of FRBs, such as the duration,
the energetics, the event rate, etc. In this review article, we will first summarize the
observations of FRBs (Section 2). We then present a detailed description of the
collision model (Section 3). In Section 4, conclusions and some brief discussion
are presented.

\section{Observed Features of FRBs}

Till the end of July 2015, only 10 FRBs are reported in the literature. They are generally
discovered through single-pulse search methods by using archive data of wide-field pulsar
surveys at the multi-beam Parkes telescope and the Arecibo telescope. FRBs are usually detected
at GHz frequency. The earliest event occurred in 2001, and the latest happened in 2014.
In Table 1, we list the key parameters of the 10 FRBs observed so far. Of these events,
only FRB 121102 is detected by the Arecibo telescope, and all other FRBs are detected by
the Parkes telescope. It thus would be very helpful if FRBs could also be detected by
other telescopes, especially by the future Chinese Five-hundred-meter Aperture Spherical
radio Telescope (FAST), which will be ready for observations in
late 2016 (Nan et al. 2011).

\begin{table}[!ht]
\caption{
Key parameters of the 10 FRBs observed till July 2015.
Note that the name of ``010125'' has been corrected from ``011025'' of Burke-Spolaor \& Bannister (2014);
The name of ``010724'' has been corrected from ``010824'' of Lorimer et al. (2007); The name of
``110626'' has been corrected from ``110627'' of Thornton et al. (2013). ``010621'' is the source of
J1852-08 of Keane et al. (2011).
The redshift is estimated by us from $z = (DM_{\rm E}({\rm cm}^{-3} \cdot {\rm pc}) - 100 )/1200 $ (Ioka 2003;
Inoue 2004), where 100 ${\rm cm}^{-3} \cdot {\rm pc}$ is the assumed DM contribution from the
host galaxy and the local environment of the FRB (Thornton et al. 2013).
The luminosity distance is calculated by assuming a flat Universe with $H_0 = 71 $ km/s/Mpc,
$\Omega_{\rm M} = 0.27$, and $\Omega_{\rm \Lambda} = 0.73$.
The references are: [1] Burke-Spolaor \& Bannister 2014;  [2] Keane \& Petroff 2015;  [3] Keane et al. 2011;
[4] Lorimer et al. 2007; [5] Thornton et al. 2013; [6] Spitler et al. 2014; [7] Ravi et al. 2015;
[8] Petroff et al. 2015.  }
\begin{center}
{\small
\begin{tabular}{cllllllllll}  
\tableline
\noalign{\smallskip}
FRB name & 010125 & 010621 & 010724 & 110220 & 110626      \\
\noalign{\smallskip}
\tableline
\noalign{\smallskip}
$l^{\rm o}$ & 356.6 & 25.4 & 301 & 50.8 & 355.8      \\
$b^{\rm o}$ & -20.0 & -4.0 & -41.8 & -54.7 & -41.7    \\
$DM$ (${\rm cm}^{-3} \cdot {\rm pc}$) & 790 & 746 & 375 & 944 & 723     \\
$DM_{\rm G}$ (${\rm cm}^{-3} \cdot {\rm pc}$) & 110 & 523 & 45 & 35 & 48    \\
$DM_{\rm E}$ (${\rm cm}^{-3} \cdot {\rm pc}$) & 680 & 223 & 330 & 909 & 675    \\
Estimated redshift $z$ & 0.48 & 0.10 & 0.19 & 0.67 & 0.48         \\
$D_{\rm L}$ (Gpc) & 2.7 & 0.45 & 0.92 & 4.0 & 2.7        \\
Duration $\tau$ (ms)  & 10.3 & 8.3 & 20 & 6.6 & 1.4         \\
$S_{\rm peak}$ (Jy) & 0.55 & 0.52 & 1.58 & 1.11 & 0.63      \\
$F_{\rm obs}$ (Jy $\cdot$ ms) & 5.6 & 4.3 & 31.5 & 7.3 & 0.9       \\
$E_{\rm radio}$ ($10^{39}$ ergs) & $\sim 1.9$ & $\sim 0.030$ & $\sim 0.99$ & $\sim 6.1$ & $\sim 0.30$   \\
Telescope & Parkes & Parkes & Parkes & Parkes & Parkes      \\
Ref. & [1,2] & [2, 3] & [2, 4] & [2, 5] & [2, 5]           \\
\noalign{\smallskip}
\tableline 
\noalign{\smallskip}
FRB name &   110703 & 120127 & 121102 & 131104 & 140514  \\
\noalign{\smallskip}
\tableline
\noalign{\smallskip}
$l^{\rm o}$ &   81.0 & 49.2 & 174.95 & 260.6 & 50.8 \\
$b^{\rm o}$ &   -59.0 & -66.2 & -0.2 & -21.9 & -54.6 \\
$DM$ (${\rm cm}^{-3} \cdot {\rm pc}$) &   1104 & 553 & 557 & 779 & 563 & \\
$DM_{\rm G}$ (${\rm cm}^{-3} \cdot {\rm pc}$) &   32 & 32 & 188 & 69 & 35 \\
$DM_{\rm E}$ (${\rm cm}^{-3} \cdot {\rm pc}$) &   1072 & 521 & 369 & 710 & 528 \\
Estimated redshift $z$ &   0.81 & 0.35 & 0.22 & 0.51 & 0.36 \\
$D_{\rm L}$ (Gpc) &   5.1 & 1.8 & 1.1 & 2.9 & 1.9 \\
Duration $\tau$ (ms)  &   3.9 & 1.2 & 3.0 & 2.0 & 2.8  \\
$S_{\rm peak}$ (Jy) &   0.45 & 0.62 & 0.4 & 2.0 & 0.47 \\
$F_{\rm obs}$ (Jy $\cdot$ ms) &   1.8 & 0.8 & 1.2 & 4.0 & 1.3 \\
$E_{\rm radio}$ ($10^{39}$ ergs) &   $\sim 2.6$ & $\sim 0.11$ & $\sim 0.055$ & $\sim 1.6$ & $\sim 0.20$ \\
Telescope &   Parkes & Parkes & Arecibo & Parkes & Parkes \\
Ref. &   [2, 5] & [2, 5] &  [2, 6] & [7] & [2, 8]  \\
\noalign{\smallskip}
\tableline 
\tableline\
\end{tabular}
}
\end{center}
\end{table}

From Table 1, we see that 6 FRBs concentrate in the Galactic latitude range of
$ 40^{\rm o} < \vert b \vert < 70^{\rm o}$, But several FRBs also occur at low and middle Galactic latitudes.
A most distinct characteristic of FRBs is their high DM values. The directly measured DMs
are in the range of 375 --- 1104 $\rm{cm}^{-3}\cdot \rm{pc}$. They are much higher than
the corresponding Galactic DMs (i.e. $DM_{\rm G}$ in Table~1). After subtracting the
contribution from our Galaxy, the DMs left (i.e. $DM_{\rm E}$ in Table~1) are still in
the range of 223 --- 1072 $\rm{cm}^{-3}\cdot \rm{pc}$, with a mean value of
602 $\rm{cm}^{-3}\cdot \rm{pc}$. DM is a measure of the column
density of free electrons along the line of sight to the source. Many authors
believe that DMs can be used to estimate the distance of these FRBs, but note that
in the practice, we are troubled by the unknown contributions from the host galaxy and
the local environment of the FRB. In Table 1, we have subtracted a value of 100
$\rm{cm}^{-3}\cdot \rm{pc}$ from $DM_{\rm E}$ and derived the
redshift for each event by using $z = (DM_{\rm E}({\rm cm}^{-3} \cdot {\rm pc}) - 100 )/1200 $ (Ioka 2003;
Inoue 2004).  As a result, we see that FRBs possibly lie in the range of
$0.1 \leq z \leq 0.81 $. The mean value of the estimated redshift is $z = 0.42$.
Consequently, the mean luminosity distance is $D_{\rm L} = 2.3$ Gpc.

FRBs are very short events. Their durations ($\tau$) are in the range of 1.2 --- 20 ms, with
the mean value being 6.0 ms. However, they are very strong radio flashes. The observed
peak flux density is $ 0.4 \, {\rm Jy} < S_{\rm peak} < 2 \, {\rm Jy}$
(mean value: $S_{\rm peak} = 0.83 \, {\rm Jy}$). As a result, the
radio fluence  is $ 0.8 \, {\rm Jy \cdot ms} < F_{\rm obs} < 31.5 \, {\rm Jy \cdot ms} $
(mean value: $F_{\rm obs} = 5.9 \, {\rm Jy \cdot ms}$). Assuming a beaming solid angle
of 1 sr, we then can give a rough estimate of the emitted radio energy, which is
$0.03 \times 10^{39} {\rm ergs} < E_{\rm radio} < 6.1 \times 10^{39} {\rm ergs}$.
Note that the mean energy released is $E_{\rm radio} = 1.4 \times 10^{39} {\rm ergs}$.

\section{Neutron Star-Asteroid Collision Model}

Various models have been proposed for FRBs. However, many of the models cannot satisfactorily
explain the basic features of FRBs, such as the short duration, the strong intensity,
the event rate, the absence of intensive emission of $\gamma$-rays, the event rate, etc. The nature
of FRBs is thus still highly controversial. We argue that FRBs may be induced by
neutron star-asteroid collision events(Geng \& Huang 2015). Our model can naturally explain the
millisecond duration of FRBs. It can also well account for various other aspects of FRBs.

\subsection{Collision process}

We assume the neutron star has a mass of $M=1.4 \, {\rm M}_{\odot}$ and its radius is $R_{\rm NS}=10^6 {\rm cm}$.
The mass and radius of the asteroid are designated as $m$ and $r_0$, respectively. For simplicity, it
is assumed to be of Fe-Ni composition, so that the density is $\rho_0 \sim 8~\rm{g~cm}^{-3}$. When
the asteroid falls toward the neutron star, it will be broken up at the tidal breakup radius
given by $R_{b} = \left(\rho_0 r_0^2 M G / s \right)^{1/3}$, where $G$ is the gravitational constant
and $s$ is the shear strength. The subsequent free fall velocities of the leading and lagging
fragments will be different (Colgate \& Petschek 1981). As a result, the disrupted material
is highly elongated. The duration of the final collision of the material with the neutron star
can be estimated as (Colgate \& Petschek 1981; Geng \& Huang 2015):
\begin{equation}
\Delta t_a = 2 r_0 \left(\frac{2 G M}{R_{b}}\right)^{-1/2}
= 1.58 \times 10^{-3} m_{18}^{4/9} s_{10}^{-1/6} \left(\frac{\rho_0}{8~\rm{g~cm}^{-3}}\right)^{-5/18}
\left(\frac{M}{1.4 M_{\odot}}\right)^{-1/3} \rm{s}.
\end{equation}
For an asteroid with $m \sim 10^{18}$ g, the collision will be completed on ms timescale. So the
short durations of FRBs can be safely ensured.

We assume a small portion ($\eta_{\rm R} \sim 10^{-2}$) of the potential energy is emitted in
radio band. Then from $f E_{\rm FRB} = \eta_{\rm R} G M m / R_{\rm NS}$, we can derive the
asteroid mass as
\begin{equation}
m = 5.4 \times 10^{17} \eta_{\rm{R},-2}^{-1} f_{-3} E_{\rm{FRB},40} R_{\rm{NS},6} M_{1.4 M_{\odot}}^{-1} {\rm g},
\end{equation}
where $f \sim 3 \times 10^{-3}$ is the estimated beaming factor of radio emission in
our model (Geng \& Huang 2015). Note that the isotropic FRB energy is $E_{\rm FRB} = 4 \pi E_{\rm radio}$.

\subsection{Coherent radiation}

Just before the final collision, the material is compressed to a dense thin sheet by the magnetic field
of the neutron star. The collision will launch a rapidly expanding plasmoid fireball from the surface,
leading to a fan of field lines filling with relativistic electrons. The thickness of the fireball
is $\Delta \approx c \tau$. The emission volume is $V_{\rm emi} \approx 4 \pi f \Delta r_{\rm emi}^2$,
where the emission radius $r_{\rm emi}$ can be determined from the balance between the plasma pressure
and the magnetic energy density. Electrons radiate coherently in patches with typical volume given
by $V_{\rm coh} = \left(4/\gamma^2\right) r_{\rm emi}^2 \times \left(c/\nu_{\rm c}\right)$,
where $\nu_{\rm c}$ is the characteristic frequency of curvature emission which naturally falls
in the radio band. The typical Lorentz factor of electron ($\gamma$) is estimated as several hundred.
The total curvature luminosity can then be derived as (Kashiyama et al. 2013)
$ L_{\rm tot} \approx \left(P_{\rm e} N_{\rm coh}^{2}\right) \times N_{\rm pat}$,
where $P_{\rm e}$ is the emission power of a single electron,
$N_{\rm coh}$ is the number of electrons in each coherent patch,
and $N_{\rm pat}$ is the number of the patches.

\subsection{X-ray afterglow}

The collision will lead to a hot region on the neutron star surface, whose temperature can
reach keV range initially. The hot region then cools down slowly, emitting X-rays in a relatively
long period. In fact, much of the released potential energy will be emitted as X-rays.
Previous studies indicate that the thermal radiation from such a hot region is
likely to decay as a power-law function of time (Lyubarsky et al. 2002). In our study, we have
calculated the X-ray afterglow of an FRB by
using $F_{\rm X} \approx \sigma T^4 R_{\rm NS}^2/d_{\rm L}^2$. For an FRB occurring
at cosmological distance, the decaying thermal emission is usually more than 12 magnitudes too
dim to be detected by current X-ray detectors.
However, note that for other FRB models such as the collapse of hypermassive neutron stars to
black holes (Zhang 2014), the energy released is much larger and X-ray afertglows due to
external shocks are detectable in some optimistic cases (Yi et al. 2014).

\subsection{Event rate}

For a typical neutron star, a strong asteroid collision event may happen every
$\sim 10^6 - 10^7$ years (Mitrofanov \& Sagdeev 1990). In the volume of $z \leq 1$,
there are about $\sim 10^9$ late-type galaxies, and each galaxy may contain $\sim 10^8$
neutron stars. As a result, we estimate that the observable FRBs caused by
collision events are $\sim 10^4 - 10^5  f_{-3}~\rm{sky}^{-1}~\rm{day}^{-1}$.
This rate is roughly consistent with that deduced from observations,
i.e., $\sim 10^{3}$ --- $10^4~\rm{sky}^{-1}~\rm{day}^{-1}$ (Thornton et al. 2013;
Keane \& Petroff 2015).

\section{Conclusions and Discussion}

As a special kind of phenomena that are characterized by very short timescale and
high intensity, the newly discovered FRBs are still highly controversial for their
bursting mechanism. We suggest that the collision between asteroids and neutron
stars can reasonably explain many of the observed features of FRBs, such as the
millisecond duration, the energetics, the event rate, etc.

Although many authors believe that FRBs are at cosmological distances,
the possibility that FRBs are actually at non-cosmological distances or even are
Galactic events still cannot be excluded yet (Katz 2014; Maoz et al. 2015; Connor et al. 2015).
Interestingly, our collision model can also explain these non-cosmological FRBs (Geng
\& Huang 2015). In this case, the asteroids involved will be much smaller.

Neutron stars/pulsars are very intriguing objects. They can act as natural laboratories
for some special conditions (Weltevrede et al. 2011; Wang et al. 2014;
Chennamangalam et al. 2015; Gao et al. 2015).
For a long time, people have expected that the collision of small bodies with
neutron stars can give birth to some kinds of X-ray bursts or some special
$\gamma$-ray/soft $\gamma$-ray bursts (e.g. Colgate \& Petschek 1981; Cordes \& Shannon 2008;
Campana et al. 2011). Recently, Huang \& Geng (2014) further suggested that such
collisions can also lead to glitches or even anti-glitches. The in-spiralling
merge of a strange quark star with a strange quark planet can even produce a
very strong gravitational wave burst, which can be used as a unique probe for
the existence of strange quark matter in the Universe (Geng et al. 2015). As
shown in this study, FRBs may also be a kind of phenomena associated with
neutron star collision events. We hope the Chinese FAST telescope (Nan et
al. 2011) would be a powerful tool to study the multidimensionality of
such collisions.

\acknowledgements This study was supported by the National Basic Research Program of China with Grant No.
2014CB845800, and by the National Natural Science Foundation of China with Grant No. 11473012.



\end{document}